# The Effect of Counterions on the Interactions of Charged Oligothiophenes

*Nicholas E. Singh-Miller\*, Damian A. Scherlis, and Nicola Marzari*

Department of Materials Science and Engineering and Institute for Soldier Nanotechnologies

Massachusetts Institute of Technology, Cambridge, MA 02139

nedward@mit.edu, damian@qi.fcen.uba.ar, marzari@mit.edu

ABSTRACT. The functionality of conjugated polymer systems often relies on oxidations or reductions, in most cases mediated by the presence of counterions. The effect that the common counterion hexafluorophosphate ($PF_6^-$) has on the intermolecular interactions between charged oligothiophenes is investigated here using *ab initio* quantum chemistry methods. Counterions are explicitly included in the simulations of oxidized oligothiophenes and in the dimerization process. Our calculations provide quantitative and qualitative insight into the intermolecular interactions in oligothiophene-counterion systems and show that the intermolecular π-stacking of oligothiophenes is not adversely affected by the presence of counterions, and that in fact oligothiophene dimerization is further stabilized by their presence.

**Introduction**

π-conjugated polymers have received considerable attention since their introduction decades ago.[1] Due to their conductive properties that can be controlled by doping, these polymers have seen applications as conductors and semiconductors,[1] with many notable examples in optoelectronics,[2,3] electronic devices,[4,5] and as components in molecular scale actuators.[6-11] It is this latter novel use of π-conjugated polymers that motivates this study, where the driving force for actuation comes from conformational changes following oxidation or reduction, as discussed in Refs. [6-11].

It has been shown previously experimentally[12] and theoretically[13-15] that when oligothiophenes are oxidized they will dimerize through π-stacking. The interaction takes place between two positively charged planar oligomers, where the strong tendency of the charged oligomers to repel each other via Coulomb interactions is overcome by π-bond hybridization and by solvation effects (such as polarization and screening of the surrounding dielectric medium[15] and surface tension of the solvent[16]). Although dimerization of conjugated oligothiophenes occurs when the oligomers are charged, it has not been common practice to explicitly describe the counterions in the electronic structure calculations.

In reality, charge compensating counterions will be necessarily present. In actuating systems the counterions can take an active or passive role in the dynamics of the system: For example, in the case of conventional polypyrrole electro-actuators, the migration of counterions into the polymer matrix upon oxidation can induce a swelling actuation mechanism,[9-11] while in other systems[6-8] the counterion may not play a definite role in the polymer behavior.

In this paper we explore the effects that the common counterion hexafluorophosphate, $PF_6^-$, has on charged oligothiophene systems, using quantum-chemistry *ab initio* techniques. In addition, we study the properties of the individual molecular species (electron affinities and ionization potentials), the charge transfer and binding properties of singly-charged oligothiophene monomers, and the binding interactions within doubly-charged, dimerized oligothiophenes.

Functionality aside, questions still remain about the overall effect that counterions have on π-conjugated systems. Few studies, experimental[17,18] or theoretical[19,20], have focused on these effects with respect to conjugated polymers. It is expected that, beyond charge compensation, the counterion may affect the structure and electrostatic properties of these π-conjugated systems. However, it should be mentioned that the role of counterions has been studied at length with respect to polyelectrolyte systems,[21] and showing e.g. how the counterions influence the structure of the polyelectrolyte chains in solution.[22]

**Computational Methods**

Structural optimizations and single-point energy calculations were carried out using Gaussian03.[23] The calculations employed density functional theory (DFT) with two different exchange-correlation functionals: the generalized gradient approximation (GGA) of Perdew-Burke-Ernzerhof[24,25] (PBE), and the hybrid functional B3LYP.[26,27] The polarizable-continuum model[28] (PCM) was used to describe systems solvated in acetonitrile (ACN, $\varepsilon=36.64$), dichloromethane (DCM, $\varepsilon=8.93$) or water ($\varepsilon=78.39$). The solvents used were chosen for a range of dielectric constants, regardless of solubility (e.g. water) of the oligomers. All calculations in Gaussian03 were performed with the 6-311+G(d,p) basis set. The basis set superposition error (BSSE) was computed using the counterpoise method,[29,30] with the exception of the solvated (PCM) cases.[31]

Potential energy surfaces were mapped using PWscf, part of the Quantum-ESPRESSO distribution,[32] also using PBE and an ultrasoft pseudopotential formulation.[33] In this case, a plane wave basis set with an energy cutoff of 30 Ryd for the wave functions and 240 Ryd for the charge density was used. The cell size is 17.99Å x 13.45Å x 13.45Å. Other relevant parameters are provided throughout the text.

**Results and Discussion**

This section is divided into three parts. First, the electronic properties of the individual hexafluorophosphate and oligothiophene ions are addressed, including electron affinities and ionization potentials. Second, the interaction between $PF_6^-$ and an oligothiophene monomer is investigated, including charge-transfer effects and binding energies. Finally, we examine the interactions between charged dimerized oligothiophenes in the presence of $PF_6^-$, and the role of the counterion in driving dimerization and structural stability. In all cases, the effects of different solvents are investigated in detail for clarity.

*Properties of the Individual Molecules*

The structure of $PF_6$ and of several oligothiophenes (from monothiophene to quarterthiophene) were determined in both the neutral and charged states (anion for $PF_6$ and cation for the oligothiophenes), together with the electron affinity of $PF_6$ and the ionization potentials of the oligothiophenes. Table 1 contains the adiabatic ionization potentials (IPs) and electron affinities (EAs) both at the B3LYP and PBE level. These values have been obtained subtracting from the total energy of the charged system in the neutral geometry the total energy of the neutral system. For the systems in solution non-electrostatic terms (cavitation, dispersion, and repulsion energies) are also included. Experimental data[34] are available for terthiophene, and reasonable agreement is found: 143 kcal/mol versus the calculated 155.4 kcal/mol (PBE) and 159.3 kcal/mol (B3LYP). A decrease in the IP accompanying the increase in oligomer length is observed in vacuum and in all solvents, which is consistent with experimental[35] values for polythiophene (~115 kcal/mol) and with a hybrid-functional calculation[36] (~126 kcal/mol). While the IPs obtained with the two levels of theory are quantitatively similar, for EAs the discrepancies are far larger. This is somewhat expected as it is well-known that local or semi-local exchange-correlation functionals underbind negative ions,[37,38] which leads to lower EAs (as it is seen here for the PBE case).

**Table 1.** Ionization potentials of the different oligothiophenes considered and electron affinity of $PF_6$ in vacuum, dichloromethane (DCM), acetonitrile (ACN), and water, calculated with B3LYP and PBE. The 6-311+g(d,p) basis set has been used; units are kcal/mol.

|  | thiophene | | bithiophene | | terthiophene | | quaterthiophene | | $PF_6$ | |
|---|---|---|---|---|---|---|---|---|---|---|
|  | B3LYP | PBE | B3LYP | PBE | B3LYP | PBE | B3LYP | PBE | B3LYP | PBE |
| Vacuum | 206.2 | 205.2 | 173.9 | 171.1 | 159.3 | 155.4 | 150.8 | 146.4 | -187.3 | -162.1 |
| DCM | 156.1 | 155.3 | 134.4 | 132.8 | 125.7 | 122.2 | 121.3 | 117.3 | -235.2 | -213.9 |
| ACN | 151.4 | 150.6 | 130.5 | 129.1 | 122.6 | 119.2 | 118.4 | 114.3 | -239.8 | -212.8 |
| Water | 150.5 | 149.7 | 130.4 | 127.3 | 121.8 | 118.4 | 117.6 | 113.6 | -240.4 | -216.4 |

*Monomer-Counterion Interactions*

A system including a positively-charged oligothiophene and a negatively-charged counterion is electrically neutral. We thus evaluated the tendency for charge transfer between a single oligothiophene and a single counterion keeping the overall system neutral. From the electron affinities and the ionization potentials found in Table 1, both B3LYP and PBE simulations should exhibit spontaneous charge transfer between the molecules for all cases considered, except for thiophenes or bithiophenes in vacuum. To illustrate this point, we studied the energy involved in this charge transfer as a function of the separation, using terthiophene and a $PF_6$ counterion. The results for 5 Å and 10 Å of separation and for all solvents are summarized in Table 2, where we report the total energy for the neutral system (terthiophene and counterion) minus the total energy of an isolated neutral terthiophene and an isolated neutral $PF_6$ (as usual for solvation studies, the non-electrostatic energy terms are also included). Charge transfer is always observed in our calculations, with the exception of the system in vacuum, for the two species 10 Å apart, and when using PBE. This failure, reflected in the Mulliken populations, is discussed in more detail later in this section.

For the solvated cases, as expected, the sum of the IP and EA are in excellent agreement with the energy associated to charge transfer at large separations (columns 3 and 2 of table 2, respectively), due to the screening of the long range Coulomb interactions by the dielectric medium, already very effective at a distance of 10 Å. In vacuum, instead, the charge-transfer energy at 10 Å is not yet converged to this same asymptotic limit due to the long-range electrostatics.

**Table 2.** Charge transfer energies for terthiophene-$PF_6$, calculated with B3LYP and PBE exchange-correlation functionals and the 6-311+g(d,p) basis set. The third column contains the values of EA+IP for the molecules at infinite separation (as obtained from EAs and IPs reported in Table 1). Units are in kcal/mol.

|        | 5Å separation | | 10Å separation | | IP+EA | |
|--------|---------------|-------|----------------|-------|--------|-------|
|        | B3LYP | PBE | B3LYP | PBE | B3LYP | PBE |
| Vacuum | -77.7 | -59.5 | -56.5 | +15.7 | -28.0 | -6.7 |
| DCM    | -108.8 | -88.8 | -112.3 | -91.6 | -109.5 | -91.7 |
| ACN    | -113.4 | -93.7 | -117.6 | -96.9 | -117.2 | -93.6 |
| Water  | -113.5 | -93.6 | -118.6 | -98.1 | -118.6 | -98.0 |

The dependence of the binding energy of the counterion to the charged terthiophene was also calculated. The counterion was moved normal to the plane of the terthiophene with the central sulfur of the terthiophene on the same axis of the phosphorous of the hexafluorophosphate. The equilibrium geometries determined for the individual charged molecules were used here. We show in Figure 1 the binding energy for this system as a function of separation, in different solvents. As expected, in vacuum there is a strong interaction between the molecules (~60 kcal/mol) due to the Coulomb attraction between them. However, in the polarizable solvents (acetonitrile and dichloromethane) the attraction is highly screened by the dielectric medium.

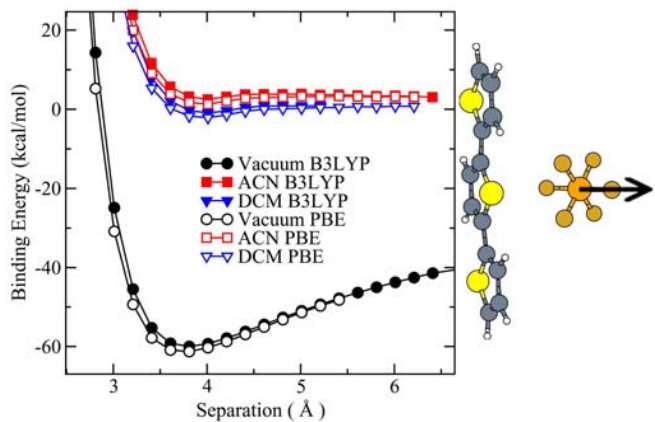

**Figure 1.** Interaction energy for the terthiophene cation and the $PF_6^-$ anion as a function of separation, in vacuum (circles), acetonitrile (squares), and dichloromethane (triangles), using B3LYP (solid) or PBE (hollow). Direction of separation is shown schematically.

We found that for separations greater than 5.4 Å convergence of the total energy was not easily achieved when studying this system in vacuum using PBE. Furthermore, when converged, the energy values were much higher than the B3LYP calculations, with a discontinuity at 5.4 Å. The Mulliken charges were investigated: when charge transfer occurs, they sum to approximately -1 on $PF_6^-$ and +1 on the terthiophene. We show in Figure 2 a plot of the sum of the Mulliken charge for $PF_6$ as a function of separation. Beyond 5.4 Å the self-consistent algorithm is not able to converge to the correct charge transfer ground state. Regardless of this failure, both functionals PBE and B3LYP were used for the remainder of this study, due to the fact that the separations between the oligomer and counterion were always close to the equilibrium separations of 3.7 to 3.8 Å.

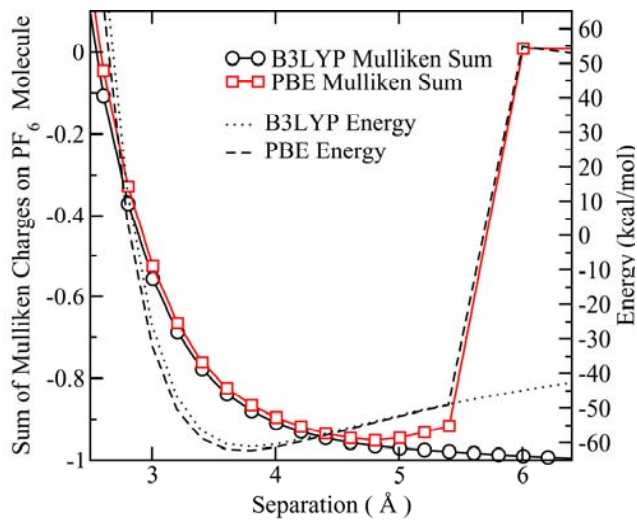

**Figure 2.** Mulliken charges for $PF_6^-$ as a function of distance from the terthiophene, for B3LYP and PBE; a value of -1 is corresponds to a full charge transfer. The binding energy is also overlaid.

Last, the potential energy surface was calculated for $PF_6^-$ in the presence of a charged terthiophene (again, an overall neutral system), using a plane-wave periodic boundary code.[32] This mapping required 63 separate ionic relaxations across the terthiophene plane using a 1 Å x 1 Å grid, as illustrated in Figure 3. The 63 points comprise one half of the system, exploiting the mirror symmetry orthogonal to the molecular plane (the 6 Å line on the horizontal axis in Figure 3). During each relaxation the sulfur atoms of the terthiophene were held fixed, the carbon atoms held in a plane (i.e. at a fixed height), and the phosphorus atom of the hexafluorophosphate was restricted in its movement to the axis perpendicular to the plane of the terthiophene (however, the fluorine atoms of the molecule were given no restriction in any direction). In this manner the counterion relaxes to its geometric minimum for each grid point without displacing the terthiophene. The potential energy surface is shown in Figure 3: most notably, it is found to be quite shallow with respect to any given point, with a difference between the global minimum and maximum of only 8.4 kcal/mol. The most favorable location for the counterion is that on the side of the terthiophene adjacent to the central sulfur atom on the chain. The location of the global minimum is further refined by full relaxations using B3LYP in both vacuum and acetonitrile: the counterion in vacuum is found slightly out of plane (0.37Å) and 3.76Å away from the central sulfur, while for acetonitrile it remains in the plane, 4.24Å away from the central sulfur atom. These minima are marked in Figure 3 and the one structural minimum found for $PF_6^-$ in vacuum will be later used to determine the potential energy surface of doubly-charged terthiophene dimers in the presence of two counterions.

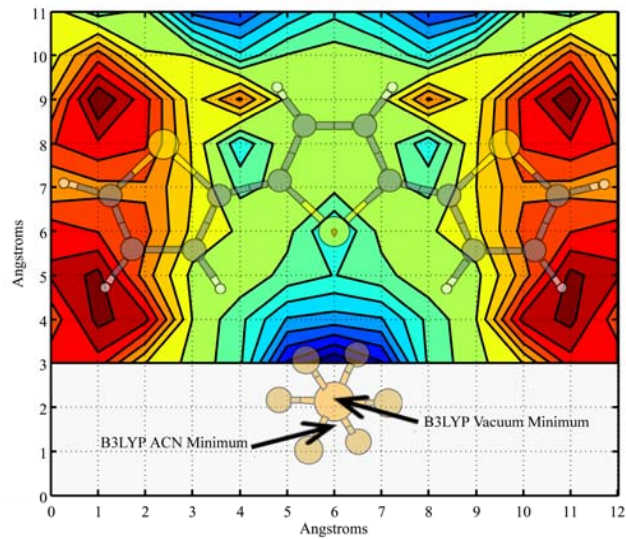

**Figure 3.** PBE potential energy surface for a $PF_6$ onto a terthiophene in vacuum. The potential energy difference between the highest point (dark red) and the lowest one (dark blue) is 8.44 kcal/mol. Also shown are the geometry minima obtained with B3LYP in both vacuum and acetonitrile (ACN); these two points were calculated using Gaussian03 and are outside the periodic box of the plane wave calculation. The molecules are schematically overlaid for visual reference.

*Dimer-Counterion Interactions*

The interaction between two charged terthiophenes in the presence of two $PF_6^-$ counterions was investigated. The dimerized assembly consists of two singly-charged terthiophenes and two counterions. As before, this is overall a neutral system, and spontaneous charge transfer within the system leads to positively-charged terthiophenes and negatively-charged hexafluorophosphate ions. The $PF_6^-$ counterions are located on opposite sides of the dimer (one in line with the central sulfur at the global minimum position, the other in line with an edge sulfur). The terthiophenes are in parallel planes separated by 3.4 Å. A sketch of this geometry is shown in Figure 4. For this geometry the binding energy of the dimer has been calculated versus the lateral displacement of the two terthiophenes, in vacuum and acetonitrile. Other geometries were also investigated and show quantitatively and qualitatively similar results. Figure 4 displays the results for the lateral displacements of oligothiophenes and counterions in vacuum and in acetonitrile. The binding energy values reported here are determined as the total energy of the dimerized system minus the energy of the two constituent systems (singly-charged terthiophene bound to one counterion). For comparison, the binding energy versus lateral displacement of the doubly-charged dimer in the absence of counterions is plotted. For this latter case, a polarizable solvent such as acetonitrile stabilizes the binding of the doubly-charged dimer, in good agreement with a previous study[15] that showed the same effect in the presence of the solvent. However it is apparent that the presence of counterions further stabilizes the dimerization process. Furthermore, even the counterions alone (i.e. without the solvent) stabilize the dimerized system, and we see that in both cases there exists energy minima in the presence of counterions that are below the limit approached at full lateral separation of the dimer. Note also that the minima in Figure 4 correspond to electronic singlet states, which suggests π-bond hybridization, whereas the maxima correspond to triplet states. PBE and B3LYP show that these minima correspond to closed-shell singlets. Previous calculations done in the absence of counterions and using several exchange-correlation functionals have shown similar results;[15] where Hartree-Fock and MP2 calculations have instead predicted for these minima an open-shell singlet configuration, with unpaired electrons of opposite spin localized on each monomer.[15] Such discrepancies between DFT and HF-based methods are attributable to the charge-density delocalization associated with the self-interaction error of the exchange-correlation functional. In spite of this disagreement regarding the nature of the singlet, the interaction energies obtained with MP2, GGA, and multiconfiguration schemes such as CASSCF turn out to be quantitatively close.[15] The maxima along the lateral displacement curve, on the other hand, are triplets corresponding to the two radical fragments exhibiting no chemical bond (note that each charged terthiophene plus the attached counterion has one unpaired electron, and so the total spin arising from two unbound counterion-terthiophene units is a triplet). The presence of the negatively-charged counterions screens the Coulomb repulsion between the positively-charged terthiophenes and ultimately leads to stable π-stacked dimers in both acetonitrile and vacuum.

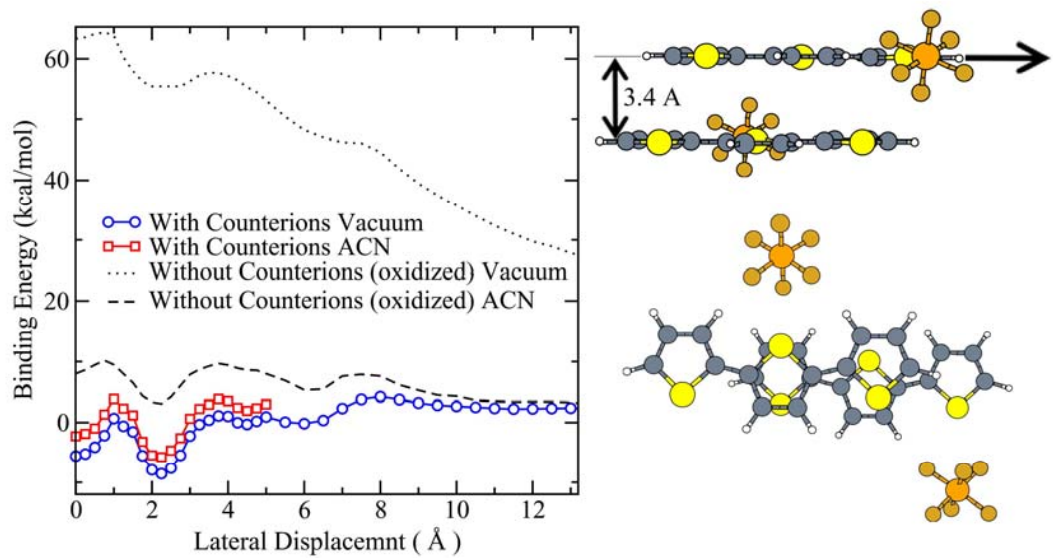

**Figure 4.** Interaction energy versus lateral displacement of the doubly-charged terthiophene dimer in the presence of PF$_6^-$ (PBE), in vacuum and in acetonitrile. The energy of the doubly-charged dimer without counterions is also plotted for comparison (B3LYP). Side and overhead views of the molecular system are shown schematically.

Finally, the energy of separation for the dimerized assembly in the presence of counterions was calculated. A laterally-displaced geometry in which the terthiophenes are in parallel planes yet displaced by 1.75Å was used for the separation-energy calculations. The geometry used, along with a plot of the binding energy for the dimer in vacuum and acetonitrile are shown in Figure 5. Again, for reference a plot of the interaction energy of the doubly-charged dimer in the absence of counterions is included in the same graph. The values reported here are the total energy of the fully dimerized system minus the total energy of the two constituent systems. Once again the presence of the polarizable solvent and the counterions stabilizes the dimerized oligothiophenes. However, for the case with no counterions in ACN the energy at the minimum is found to be positive. This is a slight deviation from the results presented in Reference 15, and can be attributed mainly to the choice of lateral displacement used here (not being the full minimum geometry) as well as the differences in the PCM and iPCM solvation methods. Regardless of this deviation, the results still show that the addition of counterions stabilizes further the π-stacked dimer.

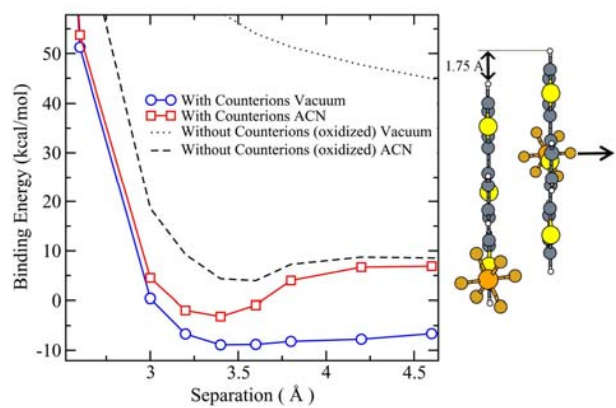

**Figure 5.** Interaction energies versus separation for the doubly-charged terthiophene dimer in vacuum and in acetonitrile (PBE). The energy of the doubly-charged dimer without counterions is also plotted for comparison (B3LYP). A side view of the molecular system is shown schematically.

**Summary and Conclusions**

Oxidized or reduced conjugated polymer systems often require the presence of counterions. We study here the effects that the common counterion hexafluorophosphate ($PF_6^-$) has on the intermolecular interactions between charged oligothiophenes, using extensive *ab initio* quantum chemistry calculations. We show that spontaneous charge transfer between hexafluorophosphate and terthiophene takes place both in vacuum (with the exception of calculations that use the PBE exchange-correlation functional at distances greater than 5.4 Å) and in solutions of dichloromethane, acetonitrile, and water. In a polar solvent the $PF_6$-terthiophene system binds weakly, with the most stable state having the counterion in plane with the terthiophene and in line with the central sulfur atom. Full charge transfer takes place, with oxidation of the terthiophene and reduction of the counterion. In the dimerization process of the charged terthiophenes, the optimal geometry has two terthiophenes staggered by 2.25 Å;

the presence of a solvent has only a very small effect on geometries. The optimal separation distance is reduced from 3.5 to 3.4 Å, and overall the system is further stabilized by ~9 kcal/mol at the minimum with respect to the case where the charged terthiophenes are in a model solvent without counterions.

**Acknowledgements**

We thank T. Swager (MIT) and I. Hunter (MIT) for sharing their expertise on polymer actuators. This research has been supported by MIT Institute for Soldier Nanotechnologies (ISN-ARO No. DAAD 19-02-D-0002); computational facilities have been provided through NSF (No. DMR-0414849) and PNNL (No. EMSL-UP-9597).

**REFERENCES**

(1) Heeger, A.J.; Kivelson, S.; Schrieffer, J.R. *Rev. Mod. Phys.* **1988**, *60*, 781-850.

(2) Gross, M.; Müller, D.C.; Nothofer, H.; Scherf, U.; Neher, D.; Bräuchle, C.; Meerholz, K., *Nature (London)* **2000**, *405*, 661-665.

(3) Moliton, A.; Hiorns, R.C. *Polym. Int.*, **2004**, *53*, 1397-1412.

(4) Pappenfus, T.; Chesterfield, R.; Frisbie, C.; Mann, K.; Casado, J.; Raff, J.; Miller, L. *J. Am. Chem. Soc.* **2002**, *124*, 4184-4185.

(5) Bredas, J-L.; Beljonne, D; Coropceanu, V.; Cornil, J. *Chem. Rev.* **2004**, *104*, 4971-5003.

(6) Yu, H.; Xu, B.; Swager, T. *J. Am. Chem. Soc.* **2003**, *125*, 1142-1143.

(7) Scherlis, D.; Marzari, N. *J. Am. Chem. Soc.* **2005**, *127*, 3207-3212.

(8) Satou, T.; Sakai, T.; Kaikawa, T.; Takimiya, K.; Otsubo, T.; Aso, Y. *Org. Lett.* **2004**, *6*, 997-1000.

(9) Gandhi, M.; Murray, P.; Spinks, G.; Wallace, G. *Synth. Met.* **1995**, *73*, 247-256.

(10) Lu, W.; Fadeev, A.G.; Qi, B.H.; Smela, E.; Mattes, B.R.; Ding, J.; Spinks, G.M.; Mazurkiewicz, J.; D. Zhou, Z.; Wallace, G.G., MacFarlane, D.R.; Forsyth, S.A.; Forsyth, M. *Science* **2002,** *297***, 983-987.**

(11) Madden, J.D.; Cush, R.A.; Kanigan, T.S.; Brenan, C.J.; Hunter, I.W. *Synth. Met.,* **2000,** *113,* 185-192.

(12) Miller, L.L.; Mann, K.R. *Acc. Chem. Res.,* **1996**, *29,* 417-423.

(13) Brocks, G. *J. Chem. Phys.* **2000**, *112*, 5353-5363.

(14) Brocks, G. *Synth. Met.* **2001**, *119*, 253-254.

(15) Scherlis, D.; Marzari, N. *J. Phys. Chem. B* **2004**, *108*, 17791-17795.

(16) Scherlis, D.A.; Fattebbert, J-L.; Marzari, N. *J. Chem. Phys.* **2006**, *124*, 194902.

(17) Nogami, Y; Pouget, J-P.; Ishiguro, T. *Synth. Met.* **1994**, *62*, 257-263.

(18) Ryu, K.S.; Jung, J.H.; Joo, J.; Chang, S.H. *J. Electrochem. Soc.* **2002**, *149(4)*, A478-A482.

(19) Mann, B.A.; Holm, C.; Kremer, K. *J. Chem. Phys.* **2005**, *122*, 154903.

(20) Pizio, O.; Bucior, K.; Patrykiejew, A.; Sokolowski, S. *J. Chem. Phys.* **2005**, *123*, 214902.

(21) Oosawa, F. Polyelectrolytes; Marcel Dekker Inc: New York, NY, 1971.

(22) Angelini, T.E.; Liang, H.; Wriggers, W.; Wong, G.C.L. *Proc. Natl. Acad. Sc. U.S.A.* **2003**, *100*, 8634-8637.

(23) Gaussian 03, Revision C.02, Frisch, M. J.; Trucks, G. W.; Schlegel, H. B.; Scuseria, G. E.; Robb, M. A.; Cheeseman, J. R.; Montgomery, Jr., J. A.; Vreven, T.; Kudin, K. N.; Burant, J. C.; Millam, J. M.; Iyengar, S. S.; Tomasi, J.; Barone, V.; Mennucci, B.; Cossi, M.; Scalmani, G.; Rega, N.; Petersson, G. A.; Nakatsuji, H.; Hada, M.; Ehara, M.; Toyota, K.; Fukuda, R.; Hasegawa, J.; Ishida, M.; Nakajima, T.; Honda, Y.; Kitao, O.; Nakai, H.; Klene, M.; Li, X.; Knox, J. E.; Hratchian, H. P.; Cross, J. B.; Bakken, V.; Adamo, C.; Jaramillo, J.; Gomperts, R.; Stratmann, R. E.; Yazyev, O.; Austin, A. J.; Cammi, R.; Pomelli, C.; Ochterski, J. W.; Ayala, P. Y.; Morokuma, K.; Voth, G. A.; Salvador, P.; Dannenberg, J. J.; Zakrzewski, V. G.; Dapprich, S.; Daniels, A. D.; Strain, M. C.; Farkas, O.; Malick, D. K.; Rabuck, A. D.; Raghavachari, K.; Foresman, J. B.; Ortiz, J. V.; Cui, Q.; Baboul, A. G.; Clifford, S.; Cioslowski, J.; Stefanov, B. B.; Liu, G.; Liashenko, A.; Piskorz, P.; Komaromi, I.; Martin, R. L.; Fox, D. J.; Keith, T.; Al-Laham,

M. A.; Peng, C. Y.; Nanayakkara, A.; Challacombe, M.; Gill, P. M. W.; Johnson, B.; Chen, W.; Wong, M. W.; Gonzalez, C.; and Pople, J. A.; Gaussian, Inc., Wallingford CT, 2004.

(24) Perdew, J.; Burke, K.; Ernzerhof, M. *Phys. Rev. Lett.* **1996**, *77*, 3865-3868.

(25) Perdew, J.; Burke, K.; Ernzerhof, M. *Phys. Rev. Lett.* **1997,** *78*, 1396.

(26) Becke, A. *J. Chem. Phys.* **1993**, *98*, 5648.

(27) Lee, C.; Yang, W.; Parr, R. *Phys. Rev. B*. **1988**, *37*, 785-789.

(28) Miertus, S.; Scrocco, E.; Tomasi, *J. Chem. Phys*. **1981**, *55*, 117.

(29) Simon, S.; Duran, M.; Dannenberg, J. *J. Chem. Phys.* **1996**, *105*, 11024.

(30) Boys, S.; Bernardi, F. *Mol. Phys.* **1970**, *19*, 553.

(31) In Gaussian03, the counterpoise (CP) method and PCM cannot be used together due to the fact that the 'dummy' atoms of CP are given a solvation shell as if they were real atoms.

(32) Baroni, S.; Dal Corso, A.; de Gironcoli, S.; Giannozzi, P.; Cavazzoni, C.; Ballabio, G.; Scandolo, S.; Chiarotti, G.; Focher, P.; Pasquarello, A.; Laasonen, K.; Trave, A.; Car, R.; Marzari, N.; Kokalj, A., http://www.quantum-espresso.org/.

(33) Vanderbilt, D. *Phys. Rev. B* **1990**, *41*, 7892.

(34) Tepavcevic, S; Wroble, A.T.; Bissen, M.; Wallace, D.J.; Choi, Y.; Hanley, L. *J. Phys. Chem. B*. **2005**, 109, 7134.

(35) Bredas, J.L. Electronic Structure of Highly Conducting Polymers. In Handbook of Conducting Polymers; Skotheim, T.A., Ed.; Dekker: New York, 1986; p 859.

(36) Salzner, U.; Pickup, P.; Poirier, R.; Lagowski, J. *J. Phys. Chem. A*. **1998**, *102*, 2572.

(37) Filippetti, A. *Phys. Rev. A*, **1998**, *57*, 914.

(38) Toher, C.; Filippetti, A.; Sanvito, S.; Burke, K. *Phys. Rev. Lett.* **2005**, *95*, 146402.